\documentclass[11pt,twocolumn,twoside]{article}
\usepackage{iccr2024}

\begin{document}

\title{Neural Graphics Primitives-based Deformable Image Registration for On-the-fly Motion Extraction} 

\author[1,2]{Xia~Li}
\author[2]{Fabian Zhang}
\author[1,3]{Muheng~Li}
\author[1,4,5]{Damien~Weber}
\author[1]{Antony~Lomax}
\author[2]{Joachim~Buhmann}
\author[1]{Ye~Zhang}

\affil[1]{Center for Proton Therapy,
          Paul Scherrer Institut, Villigen, Switzerland}
\affil[2]{Department of Computer Science,
          ETH Zurich, Zurich, Switzerland}
\affil[3]{Department of Physics,
          ETH Zurich, Zurich, Switzerland}
\affil[4]{Department of Radiation Oncology, 
          University Hospital of Zurich, Zurich, Switzerland}
\affil[5]{Department of Radiation Oncology,
          Inselspital, Bern University Hospital, University of Bern, Bern, Switzerland}

\maketitle
\thispagestyle{fancy}


\begin{customabstract}

Intra-fraction motion in radiotherapy is commonly modeled using deformable image registration (DIR). However, existing methods often struggle to balance speed and accuracy, limiting their applicability in clinical scenarios. This study introduces a novel approach that harnesses Neural Graphics Primitives (NGP) to optimize the displacement vector field (DVF). Our method leverages learned primitives, processed as splats, and interpolates within space using a shallow neural network. Uniquely, it enables self-supervised optimization at an ultra-fast speed, negating the need for pre-training on extensive datasets and allowing seamless adaptation to new cases. We validated this approach on the 4D-CT lung dataset DIR-lab, achieving a target registration error (TRE) of $1.15\pm1.15$ mm within a remarkable time of 1.77 seconds. Notably, our method also addresses the sliding boundary problem, a common challenge in conventional DIR methods.
\end{customabstract}


\section{Introduction}

Deformable image registration (DIR) is a critical technique in radiotherapy, essential for accurate motion modeling, dose accumulation~\cite{chetty2019deformable}, and image alignments. The complexity of modeling the displacement vector field (DVF) and the absence of a definitive ground truth make DIR a challenging endeavor. The evolution of DIR methodologies has transitioned from foundational techniques such as optical flow and elastic models~\cite{bajcsy1989multiresolution} to more sophisticated approaches like the demon algorithm~\cite{vercauteren2009diffeomorphic} and B-spline-based methods~\cite{rueckert2006diffeomorphic}. These advancements, including the adoption of map-based approaches, have significantly enhanced precision in capturing complex image deformations. However, they are limited by their computational time demands and the need for precise initial alignment and parameter settings.

The introduction of deep learning (DL) into DIR has been a game-changer. Notable studies like VoxelMorph~\cite{balakrishnan2019voxelmorph} have pioneered the use of unsupervised learning frameworks employing convolutional neural networks (CNNs). Despite their advances, DL-based methods still suffer from the need for extensive training datasets and face challenges in generalizing to diverse clinical scenarios. Besides, CNNs are sensitive to the input resolution (physical spacing). More recently, Implicit Neural Representations (INR)~\cite{sitzmann2020implicit} have emerged, combining the benefits of case-specific optimization with neural network efficiency. Innovations such as IDIR~\cite{wolterink2022implicit}, and ccIDIR~\cite{van2023robust} have collectively pushed the boundaries in DIR, improving processing speed and accuracy. However, they still fall short of achieving real-time performance, using at least 15 seconds for one case.

In our work, we introduce the application of Neural Graphics Primitives (NGP)~\cite{muller2022instant} for DIR, specifically adapted for 4D-CT scans in the DIR-lab dataset. By leveraging NGP for optimizing DVF, our approach combines efficiency and accuracy, utilizing learned primitives and a shallow neural network for rapid deformation field estimations. This method not only represents a significant advancement in DIR by offering quick estimations and maintaining high accuracy but also shows substantial promise for real-time clinical applications. Our preliminary findings demonstrate notable improvements in landmark accuracy, image warp accuracy, and alignment of organs-at-risk (OARs) masks, addressing challenges such as sliding boundaries. 
In this work, the capabilities of NGP for DIR have been assessed by applying it to the deformation of exhalation to inhalation CT data, extracted from 4DCT studies in the DIR-lab dataset~\cite{castillo2009framework}.

\begin{figure}[t!]
  \centering
  \includegraphics[width=1.0\linewidth]{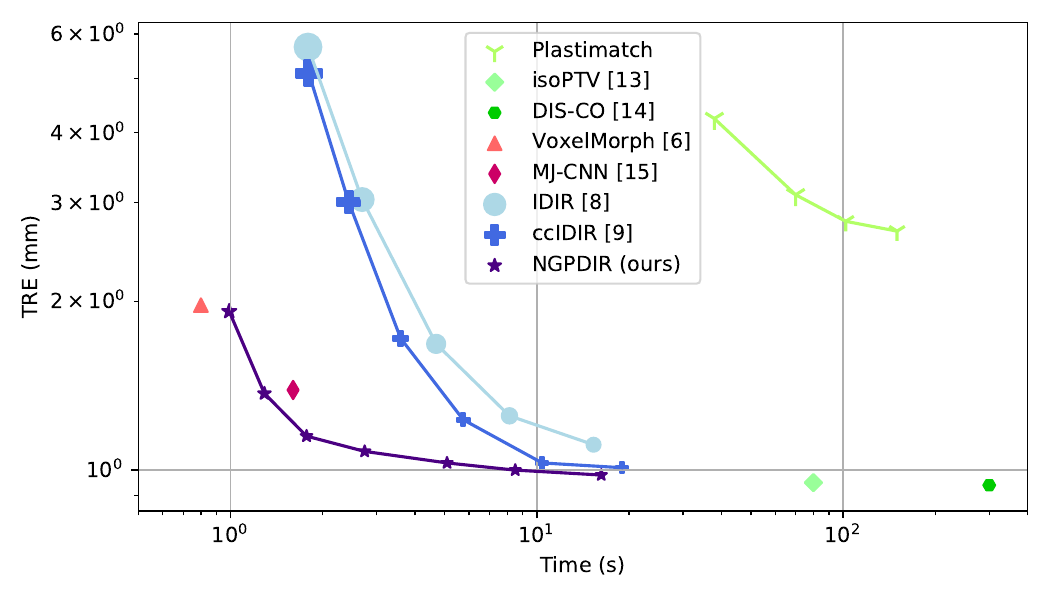}
  \vspace{-25pt}
  \caption{Landmark error, averaged over all cases, for different DIR methods. Conventional and DL-based methods are denoted in green and red, respectively, with reported values from the literature. INR/NGP based (blue) have been implemented by us and exclusively trained on the lung region for comparison.}
  \label{fig:time-tre}
  \vspace{-4mm}
\end{figure}

\begin{figure*}[t!]
  \centering
  \vspace{-15pt}
  \includegraphics[width=0.9\linewidth]{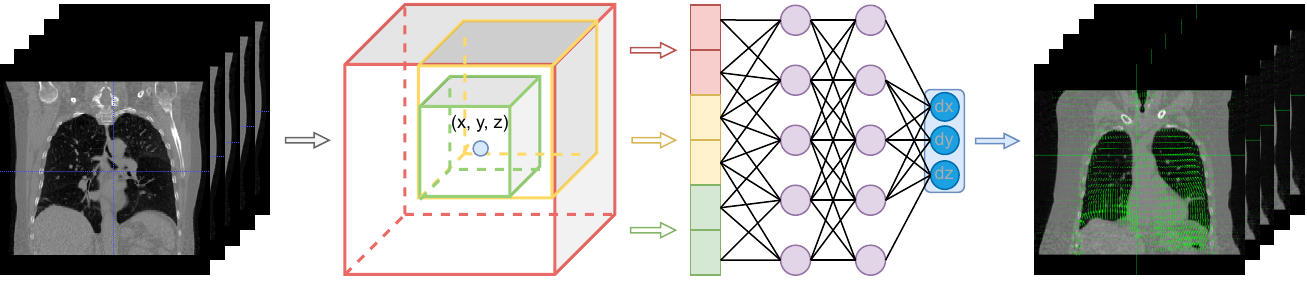}
  \caption{Schematic of the NGPDIR framework. A random point (x, y, z) is represented as the interpolations of multi-level primitives, then mapped to the displacement vector by a small network. The primitives and the network are optimized with respect to each image.}
  \vspace{-3mm}
  \label{fig:pipeline}
\end{figure*}

\section{Materials and Methods}

\subsection{Dataset}
In this study, we employed DIR-lab that comprises 4D-CT images for the radiotherapy of thoracic malignancies.
This dataset includes 10 cases, from which the exhalation and inhalation phases have been extracted for this study, with the inhalation images as the fixed images.
We preserved the original resolution of these 4D-CT scans, which varies across cases, and refrained from normalizing intensity values, as our proposed method is resolution-agnostic and value-agnostic.

\subsection{Method}

\subsubsection{INR for DIR}
The preliminary approach of our method is IDIR~\cite{wolterink2022implicit}, which adopts INR for DIR. INR fundamentally functions by mapping 3D coordinates to corresponding values or intensities by a learnable neural network, which is a crucial aspect of reconstructing and manipulating complex spatial data. In the context of DIR, INR is adapted to map these 3D coordinates to displacement vectors $(\delta x, \delta y, \delta z)$, constituting the DVF to warp the moving image (exhale phase) towards the fixed image (inhale phase). The training process of INR within DIR involves random sampling of points within the 3D space, followed by the computation of the Normalized Cross-Correlation (NCC) loss between the warped image and the fixed image, together with a Jacobian regularization. Unlike DL-based methods that require training on large-scale datasets, INR-based DIR only needs to fit one network per case, so it combines the advantages of instance optimization with neural representation.

\subsubsection{NGP for DIR}

The multi-resolution encoding in NGP employs a hierarchical hash table mechanism to represent 3D coordinates at varying levels of detail, which can be formulated as follows:

Given a 3D coordinate $\mathbf{x}=(x,y,z)$, the encoding process is defined by a set of hashing functions $\{h_1, h_2, ..., h_L\}$, where each $h_i$ corresponds to a different resolution level $i$. These functions map $\mathbf{x}$ to a series of indices that reference feature vectors within the respective hash tables:

\begin{equation}
\mathbf{f}_i = \text{HashTable}_i[h_i(\mathbf{x})],
\end{equation}

where $\mathbf{f}_i$ is the feature vector retrieved from the $i$-th hash table. The final feature representation $\mathbf{F}(\mathbf{x})$ for the coordinate $\mathbf{x}$ is then obtained by concatenating the feature vectors across all resolution levels:

\begin{equation}
\mathbf{F}(\mathbf{x}) = \bigoplus_{i=1}^{L} \mathbf{f}_i,
\end{equation}

where $\bigoplus$ denotes the concatenation operation. The feature representation $\mathbf{F}(\mathbf{x})$ is then passed through the neural network to predict the displacement vector $\mathbf{v}$:

\begin{equation}
g_{\phi}(\mathbf{F}(\mathbf{x})) \rightarrow \mathbf{v}=(\delta x, \delta y, \delta z).
\end{equation}

Through this multi-level representation, NGP can manage different scales of spatial features, which is particularly beneficial for DIR where capturing both global anatomical structure and local tissue details is crucial.

\subsubsection{Implementation details}
For fair comparisons, we have re-implemented both IDIR and ccIDIR in a shared code base with NGPDIR, which will be made public. For training, our batch size was set at $10,000$ points, randomly sampled for each iteration. Specifically, we only used points within the lung for landmark evaluation and the body for MAE calculation. For IDIR and ccIDIR, we keep their original learning rate (LR) as $1e-4$, while for our method, we employed a higher LR of $1e-2$. Our training regimen included a warmup phase for the learning rate, followed by a cosine schedule that gradually reduced it to zero towards the end of training. To analyze the relationship between training time and accuracy, we trained each method for a range of steps: [60, 125, 250, 500, 1000, 2000, 4000]. All experiments were conducted on an RTX 4090 GPU.

\begin{figure*}[t!]
  \centering
  \vspace{-15pt}
  \includegraphics[width=1.0\linewidth]{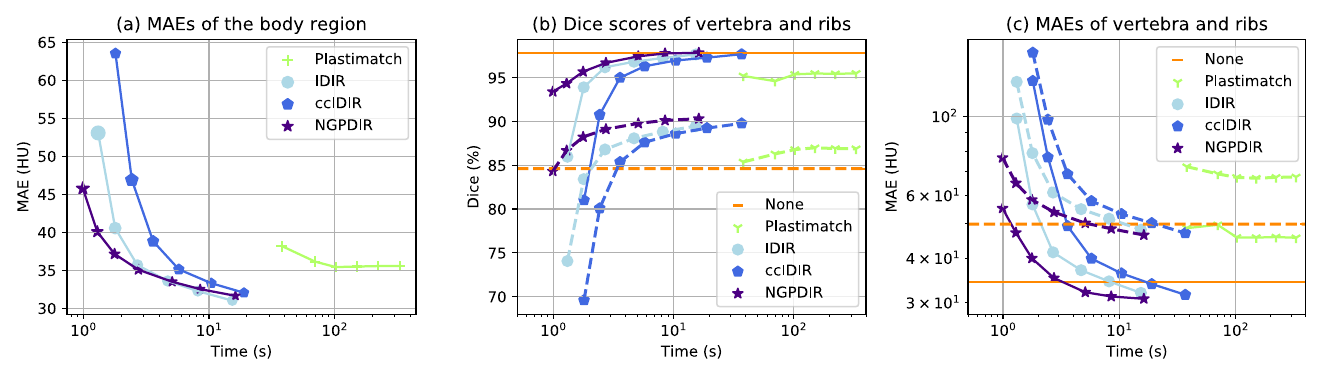}
  \vspace{-15pt}
  \caption{MAE and Dice scores for the different DIR methods averaged over all cases. Orange lines in sub-figures (b-c) represent the absence of registration as a baseline. Solid lines indicate results for the vertebra, while dashed lines correspond to ribs.}
  \label{fig:3plots}
\end{figure*}

\begin{figure*}[t!]
  \centering
  \includegraphics[width=1.0\linewidth]{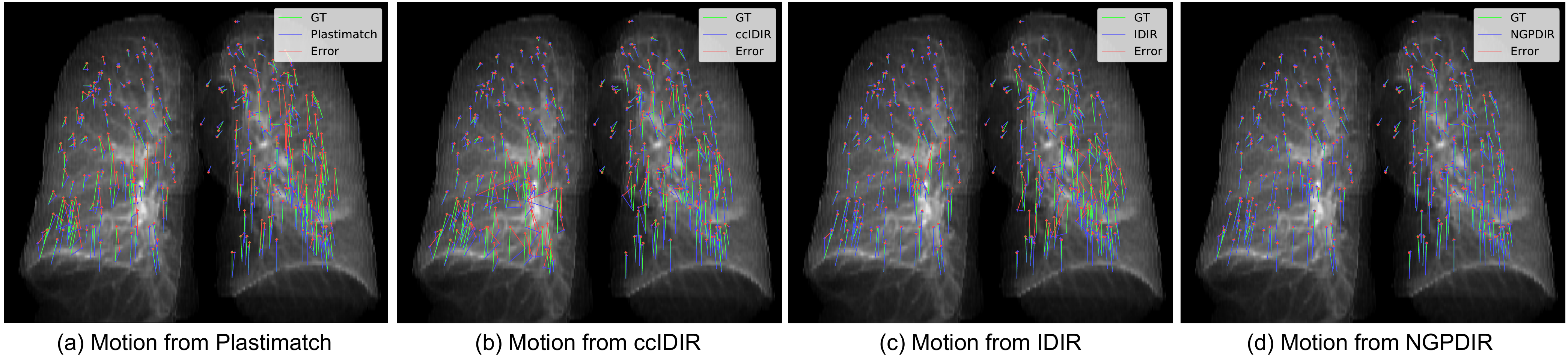}
  \vspace{-15pt}
  \caption{Visualization of landmark movement for case 8. Ground-truth motions are denoted by green arrows, predictions by blue, and errors by red. The magnitude of errors corresponds to the length of the red arrows. These projections are in the coronal plane for clarity.}
  \vspace{-5pt}
  \label{fig:marks}
\end{figure*}

\subsection{Evaluation}
Within the DIR-Lab dataset, the physician's labeled landmarks were provided as the ground truth. We further adopted Totalsegmentor~\footnote{https://totalsegmentator.com/} to get masks for sliding boundary regions like vertebras and ribs.
Evaluations were done upon landmarks using the target registration error (TRE), and also on image difference between the fixed image and the warped one using mean absolute error (MAE). Sliding boundary areas are evaluated by Dice Coefficients and MAE. Furthermore, we visually compared the predicted motion, DVF as well as warped images from different methods.

\section{Results}

Landmark prediction errors averaged over all $10$ cases of DIR-Lab are depicted in Figure~\ref{fig:time-tre}. Both IDIR, ccIDIR, NGPDIR, and Plastimatch~\cite{sharp2010plastimatch} trained on the lung region with different steps are plotted as curves. Besides, we also incorporate conventional methods (isoPTV~\cite{tancik2020fourier} and DIS-CO~\cite{ruhaak2017estimation}) and DL-based methods (VoxelMorph~\cite{balakrishnan2019voxelmorph} and MJ-CNN~\cite{jiang2020multi}) for comparisons. 
From the figure, conventional methods have the lowest landmark errors, while DL-based methods exceed in speed. Our NGPDIR achieved the best trade-off by far between speed and accuracy. 

Other metrics are plotted in Figure~\ref{fig:3plots}, where the same trend can be observed.
Notably, the conventional B-spline-based method (Plastimatch) cannot handle the sliding boundary regions well, achieving even worse results than the baseline (no registration), while NGPDIR quickly exceeds the baseline after training for only $1.2$ seconds. Besides, ccIDIR displayed slower convergence compared to IDIR when trained on the entire body, contravening the trend in Figure~\ref{fig:time-tre}.

\begin{figure*}[t!]
  \centering
  \includegraphics[width=1.0\linewidth]{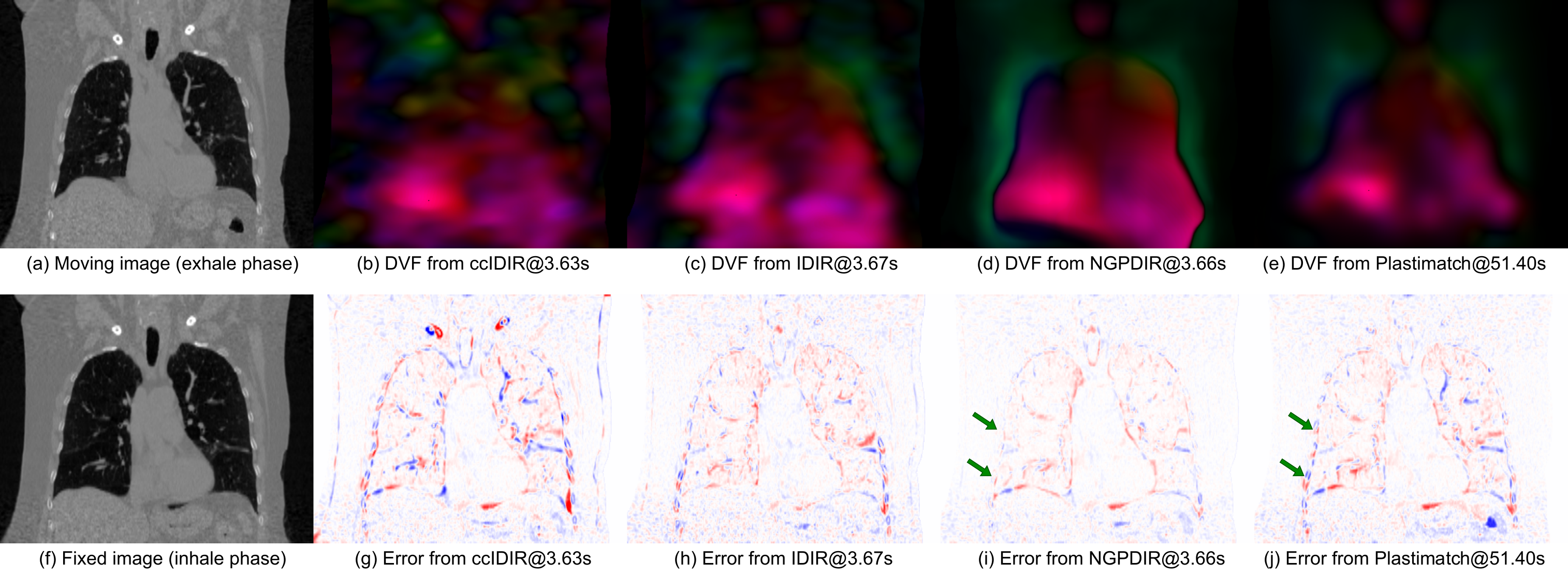}
  \vspace{-20pt}
  \caption{Example DVFs and error maps for case 8. ccIDIR, IDIR, and NGPDIR are compared with similar training times (around 3.65s). DVFs are color-coded for directionality, with the intensity reflecting the magnitude. Green arrows point to ribs (sliding boundaries).}
  \vspace{-1mm}
  \label{fig:comp}
\end{figure*}

The landmark motion visualizations in Figure~\ref{fig:marks} further depict NGPDIR's precision, with the shortest error vectors. Moreover, NGPDIR's DVF is notably smoother and more accurate, particularly in the most challenging case (Case No. 8), highlighted in Figure~\ref{fig:comp}. Meanwhile, it surpasses all the others in the error map, outperforming Plastimatch significantly around the ribs (pointed by green arrows).

\section{Discussion}

In our pursuit of advancing DIR, we draw inspiration from the dynamic realm of 3D reconstruction techniques. Notably, IDIR has leveraged the foundational principles of INR and we harness the computational efficiency derived from Instant-NGP. Given that both INR and NGP are per-point query methodologies, their natural extension into DIR estimation involves querying all voxels within a volume. Further, this can be extended volume grids, bestowing our approach with the valuable property of resolution independence. This unique trait empowers our approach with consistent performance across images of varying sizes, distinguishing it from conventional DL-based methods, which can be sensitive to input resolution and size.

Our comparative analysis reveals the remarkable versatility and efficiency of this approach when compared against both Deep Learning (DL)-based and conventional DIR methods. By effectively eliminating the need for extensive datasets while excelling in per-case optimization, our approach distinguishes itself from DL-based methods. Furthermore, our utilization of neural networks and hash encoding techniques expedites convergence, presenting an appealing alternative to conventional DIR methods.

Turning our attention to the broader implications of our work, we recognize that the current limitation of implementing INR/NGP for DIR lies in their need for case-by-case fitting. As we look forward, there is exciting potential for further generalization through meta-learning. This approach has the capacity to learn a general prior for specific human regions, followed by case-specific information fitting. 

While we propose a fast and efficient DIR approach, it's important to acknowledge the potential bottlenecks that may hinder its real-time application. Challenges persist in the realms of image acquisition hardware and image reconstruction algorithms. Improved acquisition hardware and faster reconstruction algorithms are needed to fully realize the potential of our accelerated DIR method. Additionally, the evolving landscape of 3D reconstruction technologies holds promise, offering solutions to address the later challenges. Beyond real-time image guidance, the rapid speed of our DIR approach opens the door to a range of applications in radiotherapy, including adaptive treatment planning, organ motion tracking, and dose optimization to name a few.
\section{Conclusion}

In this study, we have successfully integrated NGP into DIR, a novel contribution that significantly enhances the accuracy and efficiency of medical image alignment as demonstrated on the DIR-lab dataset. The NGPDIR framework exhibits robust performance across various metrics, particularly in landmark alignment precision and the accommodation of anatomical sliding boundaries. This advancement not only propels the DIR field forward but also opens new avenues for real-time clinical applications, potentially transforming patient care with its rapid, reliable imaging capabilities.

\printbibliography

\end{document}